\documentstyle[12pt]{article}
\newskip\humongous \humongous=0pt plus 1000pt minus 1000pt
  \newif\ifdtup

 \def\half{\mbox{\small $\frac{1}{2}$}}

\def\ltap{\raisebox{-.4ex}{\rlap{$\sim$}} \raisebox{.4ex}{$<$}}

\def\frac#1#2{ {{#1} \over {#2} }}

\def\abar{{\bar \alpha_S}}
\def\abn{{{\bar \alpha_S} \over \omega}}
\def\as{\alpha_S}

\def\msbar{\overline{\mbox {\rm MS}}}

\def\ga{\gamma}

\def\w{\omega}

\def\rt){\right)}
\def\lt({\left(}
\def\rq]{\right]}
\def\lq[{\left[}

\def\figcap{\section*{Figure Captions\markboth
        {FIGURECAPTIONS}{FIGURECAPTIONS}}\list
        {Figure \arabic{enumi}:\hfill}{\settowidth\labelwidth{Figure
999:}
        \leftmargin\labelwidth
        \advance\leftmargin\labelsep\usecounter{enumi}}}
 \relax

\def\np#1#2#3{Nucl.\ Phys.\ B#1 (19#3) #2}
\def\pl#1#2#3{Phys.\ Lett.\ #1B (19#3) #2}
\def\pr#1#2#3{Phys.\ Rev.\ D #1 (19#3) #2}
\def\prep#1#2#3{Phys.\ Rep.\ #1 (19#3) #2}

\def\sj#1#2#3{Sov.\ J.\ Nucl.\ Phys.\ #1 (19#3) #2}
\def\zp#1#2#3{Zeit.\ Phys.\ C#1 (19#3) #2}
\catcode`\@=11

\def\ps@headings{\def\@oddfoot{}\def\@evenfoot{}
\def\@oddhead{\hbox{}\hfill --\thepage{}-- \hfill}
\def\@evenhead{\@oddhead}
\def\subsectionmark##1{\markboth{##1}{}}
}

\ps@headings

\catcode`\@=12

\relax

\relax

\begin{document}

\begin{titlepage}
\renewcommand{\thefootnote}{\fnsymbol{footnote}}
\begin{flushright}
     Cavendish-HEP-94/18 \\
     Fermilab-PUB-95/006-T \\
     January 1995
     \end{flushright}
\vspace*{5mm}
\begin{center}
{\Large \bf \boldmath QCD Scaling Violation at Small $x$}\\
\vspace*{1cm}

        \par \vskip 5mm \noindent
        {\bf R. K. Ellis}\\
        \par \vskip 3mm \noindent
        Fermi National Accelerator Laboratory\\
        P. O. Box 500\\
        Batavia, IL 60510, USA\\
        \par \vskip 5mm \noindent
        {{\bf F. Hautmann$^*$} and {\bf B. R. Webber\footnote{Research
        supported in part by the UK Particle Physics and Astronomy
        Research Council and the EC Programme ``Human Capital and
        Mobility", Network ``Physics at High Energy Colliders", contract
        CHRX-CT93-0357 (DG 12 COMA).}}}\\
        \par \vskip 3mm \noindent
        Cavendish Laboratory\\
        Department of Physics, University of Cambridge\\
        Madingley Road, Cambridge CB3 0HE, UK\\

\par \vskip 1cm

\end{center}
\vspace*{1cm}

\begin{center} {\large \bf Abstract} \end{center}
\begin{quote}
We investigate the evolution of parton densities
at small
values of
the momentum fraction, $x$,
by including resummed anomalous dimensions
in the renormalization group equations.
The resummation takes into account
the leading-logarithmic contributions
$(\as \ln x)^k$  given by the BFKL equation
and  the next-to-leading-logarithmic
corrections from quark evolution.
We present numerical results for the parton densities
and the deep inelastic structure function $F_2$.
\end{quote}
\vspace*{2cm}
\end{titlepage}

\renewcommand{\thefootnote}{\fnsymbol{footnote}}

\renewcommand{\thefootnote}{\fnsymbol{footnote}}

\noindent {\bf \boldmath 1. Small-$x$ processes in QCD perturbation
theory }
\vskip 0.1 true cm
Perturbative QCD provides a good description of
hadronic collisions at large values of the centre-of-mass energy
$\sqrt S$ and the transferred momentum $p_t$, in the region
$ \Lambda_{QCD}^2 \ll p_t^2 \sim S$ [\ref{Stirl}].
This description is based on the renormalization group
analysis of physical cross sections in terms of
universal parton densities and process-dependent
coefficient functions.  The parton densities $f_a(x,\mu^2)$
($a=q_i,{\bar q}_i,g, \;i=1,\dots,N_f, \; N_f$ being the number of
active flavours) are functions of the parton
momentum
fraction $x$ and the hard scale $\mu^2\sim p_t^2$.
In leading-twist order, they satisfy the evolution equations
\begin{equation}
\label{apeq}
\frac{df_a(\w,\mu^2)}{d\ln \mu^2} =
\sum_b \ga_{ab}(\w,\as(\mu^2))\, f_b(\w,\mu^2)
\;\;,
\end{equation}
where $\w$ is the moment variable conjugate to $x$,
\begin{equation}
\label{mom}
f_a(\w,\mu^2) = \int_0^1 dx\,x^\w\,f_a(x,\mu^2)\;,
\end{equation}
the evolution kernels $\gamma_{ab}(\w,\as)$ are the
appropriate anomalous dimensions, and the running coupling
$\as$ is determined by
\begin{equation}
\label{running}
\frac{d\,\as(\mu^2)}{d\ln \mu^2} = -b_0 \as^2 - b_1 \as^3 + {\cal
O}(\as^4)
\;\;\;\;
(b_0={{33-2\,N_f} \over {12 \pi}} \;,\;  b_1={{153-19\,N_f} \over {24
\pi^2}})
\;\;.
\end{equation}
Scaling violation is systematically taken into account to
leading, next-to-leading, etc., accuracy in $\ln\mu^2$ by expanding
the anomalous dimensions to fixed order in $\as$,
\begin{equation}
\label{pert2}
\gamma_{ab}(\w,\as)=
\gamma_{ab}^{(1)}(\w) \, \as +
\gamma_{ab}^{(2)}(\w) \, \as^2 + {\cal O} (\as^3) \;\;\,,
\end{equation}
and using the corresponding expansions of the
coefficient functions.

At present and future high-energy colliders,
a new
perturbative
regime opens up, characterized by small values of
$x$ (when
$\Lambda^2_{QCD}\ll p_t^2 \ll S)$. In particular,
plentiful data on deep inelastic lepton scattering
at low $x$ are coming from the HERA $ep$ collider
and fixed-target experiments [\ref{felt}].
In the small-$x$ regime, novel dynamical effects enter [\ref{revx}]
as a consequence of the presence of the two very different
large scales $p_t^2$ and $S$. The new effects are related to
the emission of many gluons, widely separated in rapidity
and disordered in transverse momentum.
They lead to logarithmic enhancements $(\as \ln x)^k$
in higher-loop contributions to the
perturbative expansions of physical cross sections.
Correspondingly, the anomalous dimensions contain
singularities at $\w \to 0$
to all orders in perturbation theory, as follows
\begin{equation}
\gamma_{ab}(\w,\as) = \sum^{\infty}_{k=1}
\left(\frac{\as}{\w}\right)^{k} A^{(k)}_{ab} +
\sum^{\infty}_{k=0}  \as
\left(\frac{\as}{\w}\right)^{k} B^{(k)}_{ab} +
{\cal O}(\as^2(\as/\w)^k)
\;\;
\label{gam}
\end{equation}
(analogous singularities show up in the coefficient functions as well).
These contributions may spoil the convergence of the perturbative
expansion at small $x$. Indeed, numerical studies in fixed-order
perturbation theory have already found this instability in the $x$
range accessible at HERA energies [\ref{EKL}].
The singular terms should be summed to all orders in
perturbation theory to obtain reliable predictions.

Eq.~(\ref{gam}) may be taken to define
the logarithmic hierarchy at small $x$. We shall refer
to the first tower of terms $A^{(k)}$ as the
leading-logarithmic (L$(x)\,$) series, the second tower $B^{(k)}$
as the next-to-leading-logarithmic (NL$(x)\,$) series, and so on.
To L($x$) accuracy,
small-$x$ contributions are resummed by the BFKL equation [\ref{BFKL}].
Beyond the leading logarithms, the study of small-$x$ phenomena
has been the object of much recent effort, including calculational
programmes to compute ${\cal O}(\as)$-corrections
to the kernel of the BFKL equation [\ref{FL},\ref{white}],
as well as phenomenological models incorporating certain classes
of sub-leading effects into that equation [\ref{oldkwie}-\ref{Dur}].
At this level, there is an interplay between small-$x$ dynamics
and the mass singularities associated with the
low transverse momentum
region. High-energy factorization [\ref{HEF}] provides a
consistent way of combining small-$x$ resummation
and leading-twist factorization of mass singularities.
Next-to-leading
results
have been obtained
in Refs.~[\ref{CHlett},\ref{CH}] using this method.

The purpose of the present work is to investigate
quantitatively
the impact of small-$x$ dynamics
on the evolution of parton densities, and,
in particular,
on scaling violation in deep inelastic scattering. To do
this, we use resummed evolution kernels, and make
predictions for the parton densities by solving
the resulting renormalization group equations.
Note that
higher-twist effects, such as
unitarity corrections at asymptotic energies
[\ref{levrysk},\ref{uni}],
are not dealt with in this treatment. Also,
as no attempt is made
to extend the BFKL equation to sub-leading orders,
our results are not absolute predictions, but depend
upon
non-perturbative input
densities.

The system of evolution equations (\ref{apeq}) can be
separated into flavour singlet and non-singlet components.
It can be shown that the  latter are not subject to
small-$x$ corrections of the type (\ref{gam}). In what
follows it will always be understood that the non-singlet
equations are treated in the standard way using
the expansion (\ref{pert2}) up to two-loop accuracy
[\ref{GLAP},\ref{2loop}]. The enhanced small-$x$
contributions are all associated with the singlet sector.
The  singlet evolution equations can be written in the
matrix form
\begin{equation}
\label{sieq}
   {d \over {d \ln \mu^2}} \,
\left(\begin{array}{c}
{f_{S}}\\
{f_{g}}\\
\end{array}\right)\;
=
\left(\begin{array}{cc}
{
\ga_{S S}
} & { \ga_{S g }}\\
{ \ga_{g S }} & { \ga_{g g }}\\
\end{array}\right) \;\,
\left(\begin{array}{c}
{f_{S}}\\
{f_{g}}\\
\end{array}\right)\;
\;\;\;,
\end{equation}
where the quark singlet density is defined as
$f_S = \sum_{i=1}^{N_f} (f_{q_i}+f_{{\bar q}_i})$.
In the next Section we summarize the main resummed results on
the matrix  kernel in Eq.~(\ref{sieq}).  In Sect.~3 we match
them
with fixed-order contributions of the type (\ref{pert2}),
find the corresponding
solutions of the evolution equations, and
determine
predictions for the deep inelastic structure function $F_2$.

In this Letter we concentrate on presenting the main
results of our analysis.  A more detailed discussion
will be reported elsewhere [\ref{EHWprep}].

\vskip 0.4 true cm

\noindent {\bf \boldmath 2. Anomalous dimensions at small $x$ }
\vskip 0.1 true cm

At leading logarithmic (L($x$)) level,
QCD evolution is only fed by the gluon channel,
and the resummation of the ${\cal O}(\as/\w)^k$
contributions to the gluon anomalous dimensions is
accomplished by the BFKL equation [\ref{BFKL}].
The resulting structure of the singlet anomalous
dimension matrix is
\begin{equation}
\label{gad}
{\mbox{\boldmath $\gamma \!$\unboldmath}\,}_L =
\left(\begin{array}{cc}
0 & 0 \\ {C_F \over C_A} \ga_L(\w) & \ga_L(\w) \\
\end{array}\right) \;\,
+ {\cal O}(\as (\as/\w )^k) \;\;,
\end{equation}
where the BFKL anomalous dimension $\ga_L(\w)$ is
determined by the implicit equation
\begin{equation}
\label{andim}
\w = \abar\, \chi (\ga_L)
\;, \;\;\; \abar\equiv C_A\as/\pi \;,\;\;\;
\chi (\ga) \equiv 2 \psi(1) - \psi(\ga) - \psi(1-\ga) \;\;,
\end{equation}
$\psi$ being the Euler $\psi$-function.
At large values of $\w/\abar$,
the solution of Eq.~(\ref{andim}) can be obtained
by expanding the characteristic function $\chi(\ga)$
for small values of $\ga$.  The first terms of the
power series expansion in $\as$ for $\ga_L(\w)$ are
\begin{equation}
\label{gaper}
\ga_L(\w) = \abn + 2.40 \left( \abn \right)^4
+ {\cal O}\left( \abn \right)^5
\;\;.
\end{equation}
In the  region of small  $x$, however, smaller and smaller values of
$\w/\abar$ are probed, and a full inversion of Eq.~(\ref{andim})
has to be performed. This procedure leads to a
many-valued function of the complex moment variable $\w$.
The perturbative branch of $\ga_L(\w)$ is defined by requiring
it to match the expansion (\ref{gaper}) at large $\w$.
The full solution, found numerically using Newton's method,
 is shown in Fig.~1.  We see that the all-order resummation
of the perturbative $\w$-poles builds up a branch-point singularity
at the value $\w_L = 4 \ln 2\,\abar \simeq 2.77\,\abar $ in the
complex $\w$-plane. Correspondingly, as $\w$ decreases through $\w_L$
the BFKL anomalous dimension $\ga_L(\w)$ departs from the
fixed-order behaviour: its real part rises quickly until it
saturates at the  value $\ga_L=\half$, and its imaginary part
develops a discontinuity along the real axis. In $x$-space,
this singularity of the anomalous dimension gives rise to
a power-like behaviour $x^{-\w_L}$ of the gluon density at
asymptotically small values of $x$ [\ref{revx}]. The only
other singularities of $\ga_L(\w)$ on the perturbative sheet
are a complex conjugate pair of branch-points at
$\w = (-1.41\pm 1.97\,i)\abar$ [\ref{EHWprep}].
We avoid these singularities when choosing a contour
for the inversion of Eq.~(\ref{mom}).

At the next-to-leading logarithmic (NL($x$))
level, quarks start to contribute to the evolution
at small $x$ on the same footing as gluons. The NL($x$)
contribution to the singlet anomalous dimension matrix
can be written as
\begin{equation}
\label{paramdis}
{\mbox{\boldmath $\gamma \!$\unboldmath}}_{NL} =
\left(\begin{array}{cc}
{C_F \over C_A} \left[\ga_{NL}(\w)
- \frac{2\as}{3\pi}T_f \right] & \ga_{NL}(\w) \\
\ga_\delta & \ga_\eta \\
\end{array}\right)
+ {\cal O}(\as^2(\as/\w)^k)
\end{equation}
($T_f=T_R N_f = \half N_f$), where the corrections
$\gamma_\delta$, $\gamma_\eta$
to the gluon anomalous dimensions are unknown at present, while
the contributions $\ga_{NL}(\w)$ to the quark anomalous
dimensions have recently been computed
[\ref{CHlett},\ref{CH}].
In
the DIS  factorization scheme [\ref{AEM}]
$\ga_{NL}(\w)$ is given explicitly in terms
of the BFKL anomalous dimension $\ga_L(\w)$ as\footnote{Formulae are also
given in Refs.[\ref{CHlett},\ref{CH}] for the
 $\msbar$ scheme [\ref{msbar}]. Results for both schemes will be
presented in Ref.[\ref{EHWprep}].}
\begin{eqnarray}
\label{res}
\ga_{NL}^{DIS}(\w,\as) &=& \frac{\as}{\pi} \,T_f \,
\frac{2+3\ga_L-3\ga_L^2}
{3-2\ga_L} \;\frac{\Gamma^3(1-\ga_L) \,
\Gamma^3(1+\ga_L)}{\Gamma(2+2\ga_L)
\,\Gamma(2-2\ga_L)} \, R(\ga_L) \\
\mbox{where}\;\;R(\ga) &\equiv& \left\{ \frac{\Gamma(1-\ga) \;
\chi(\ga)}{\Gamma(1+\ga)
\;[-\ga \,\chi^{\prime}(\ga)]} \right\}^{\frac{1}{2}}
\exp \left\{ \ga\,\psi(1) + \int_0^{\ga} d\zeta
\;\frac{\psi^{\prime}(1) - \psi^{\prime}(1-\zeta)}{\chi(\zeta)}
\right\}\;,\nonumber
\end{eqnarray}
$\Gamma$ being the Euler $\Gamma$-function, and
$\chi$, $\chi^{\prime}$ the characteristic function in
Eq.~(\ref{andim}) and its first derivative, respectively.
Eq.~(\ref{res}) introduces no singularities in the $\w$-plane
to the right of
$\w_L$.
As in the BFKL case,
at moderate values of $\w$, $|\w| > \w_L$, an accurate
representation of the quark anomalous dimension
can be obtained by taking the first few perturbative
contributions to Eq.~(\ref{res}),
\begin{equation}
\label{qadper}
\ga_{NL}^{DIS}
= \frac{2\as}{3\pi} \,T_f \,
\left\{1+2.17 \abn +2.30 \left(\abn \right)^2 +8.27
\left(\abn \right)^3
+ {\cal O} \left(\abn \right)^4
\right\} \;\;.
\end{equation}
Note that, unlike the expansion (\ref{gaper}) for $\ga_L$,
all the perturbative coefficients in Eq.~(\ref{qadper}) are
non-vanishing, suggesting an earlier departure from the
lowest-order behaviour in the quark sector [\ref{CHlett},\ref{EKL}].
At very small values of $x$, the fully resummed result
(\ref{res}) has to be used.
As $\ga_L(\w)$ approaches its saturation value at $\ga=\half$,
the quark anomalous dimension (\ref{res}) increases rapidly,
with the behaviour $1/(1-2 \ga_L)^{1/2}$, leading to
stronger scaling violation.

\vskip 0.4 true cm

\noindent {\bf 3. Parton evolution }
\vskip 0.1 true cm

We now consider the evolution as determined by the
resummed kernels
described in Sect.~2.
We match the BFKL anomalous dimensions
(\ref{gad}) and the next-to-leading-logarithmic
results
(\ref{paramdis}) with the full one- [\ref{GLAP}]
and two-loop [\ref{2loop}] expressions,
subtracting the resummed terms to avoid
double counting, as follows
\begin{equation}
\label{matchnlo}
\ga_{ab}^{NLO}(\w,\as)=
\ga_{ab}^{(1)}(\w) \, \as +
\ga_{ab}^{(2)}(\w) \, \as^2 +
\sum^{\infty}_{k=3}
\left(\frac{\as}{\w}\right)^{k} A^{(k)}_{ab} +
\sum^{\infty}_{k=2}  \as
\left(\frac{\as}{\w}\right)^{k} B^{(k)}_{ab} \;\;.
\end{equation}
We evaluate the running coupling at two-loop level from
Eq.~(\ref{running}).
Since resummed expressions for the next-to-leading gluonic
contributions $\ga_\delta, \ga_\eta$ in Eq.~(\ref{paramdis})
are not yet available, we set them equal to their two-loop
values. We call the resulting expressions the
next-to-leading quark (NLQ($x$)) approximation.
For deep inelastic lepton scattering, the quark
anomalous dimensions are more relevant, because
the structure functions couple directly
to quarks, whereas they couple to gluons via an
${\cal O}(\as)$-suppressed coefficient function.

One should note that small-$x$ resummation does not preserve
the momentum sum rules  $\sum_a \ga_{ab}(\w=1,\as) = 0$,
because it neglects all terms other than those singular at
$\w=0$. Momentum conservation can be restored to all-loop order
by introducing contributions to the anomalous dimensions
which are subdominant at $\w\to 0$ (leaving the leading
and next-to-leading logarithmic series unchanged) but
enforce the correct behaviour at $\w\to 1$. We do this
by multiplying the resummed anomalous dimensions
by an overall factor of $(1-\w)$. To test the sensitivity
of the results to this prescription, we shall also show
results using a `harder' model of momentum conservation,
in which we simply subtract from each resummed expression
a constant equal to its value at $\w=1$.

We have solved the evolution equations assuming various
alternative
sets of parton distributions
as input conditions at scale $Q_0^2=4$ GeV$^2$.
In this Letter we limit ourselves to
the case of the flat MRSD$0^{\prime}$
 [\ref{MRS}]
input distributions.
Flat parton distributions at low $Q^2$ are suggested by the
fixed-target data of the E665 Collaboration [\ref{E665}],
and it is of particular interest to see whether they could
be consistent with the steep rise at small $x$ seen at
higher $Q^2$ at HERA [\ref{DATA}].  Data at low $Q^2$
and small $x$ from HERA should also be available soon.
We refer to Ref.~[\ref{EHWprep}] for a more complete
discussion including different inputs.

The resulting DIS-scheme singlet densities are shown in
Fig.~2 as functions of $x$ for different values of the
hard scale $Q^2$. Comparison with the fixed-order prediction
shows that the effects of the L($x$) BFKL resummation
set in at $x$-values of the order of $10^{-3}$.
They increase as $x$ becomes smaller, and are still
moderate in the range ($x \ltap 10^{-4}$) accessible at present
colliders. This is in qualitative agreement with earlier
estimates based on models which seek to extend the BFKL equation to
include a running coupling [\ref{oldkwie},\ref{bryanpino}].

We see on the other hand that, in the range of $x$ and $Q^2$ values
shown, the resummation effects from quark evolution are quite sizeable,
and tend to be larger than those in the gluon channel, in spite of
their being formally sub-leading. In particular,
they set in quite early in $x$, around values of $x \sim 10^{-2}$.
This is in agreement with expectations based on comparing
the perturbative expansions (\ref{gaper}) and (\ref{qadper}).

It should be noted, however, that the impact of
momentum-conserving corrections is model-dependent.
The difference between the results obtained using our preferred
procedure and the harder model provides an indication
of the theoretical uncertainty on the
present predictions at small $x$
due to unknown sub-dominant contributions.

The results
described above
can be used
directly
to obtain resummed predictions for the
deep inelastic structure function $F_2$.
The NLQ(x) result,
again for the case of the flat MRSD$0^{\prime}$ input distributions,
is the solid curve reported in Fig.~3.
The 1993 ZEUS data [\ref{zeus93}] are also shown for reference purposes.
We see that
a steep rise of $F_2$ at small $x$
is already obtained in the HERA range,
starting from flat input distributions,
as a consequence of perturbative resummation.
As in Fig.~2,
the next-to-leading corrections
in the quark sector have a far more dramatic impact
than the BFKL contributions.

\vskip 0.4 true cm

\noindent {\bf 4. Summary }
\vskip 0.1 true cm

We have
performed a study of
QCD dynamics
at small $x$
by incorporating resummed anomalous dimensions
into the renormalization group evolution equations.
We have used the resummed equations to
determine
predictions for the singlet parton densities and
for the deep inelastic structure function $F_2$.

Our results suggest that the fixed-order evolution of
parton densities which are flat at small $x$ for
$Q^2_0 \sim 4$ GeV$^2$ is a poor approximation.
Effects from leading and next-to-leading resummation at small $x$
are already sizeable in the HERA region, and
give rise to a steep increase of the
structure
function $F_2$.
In particular,
 the dominant contribution to this behaviour comes
from the next-to-leading corrections in the quark channel.

At the moment, the resummed predictions are
still subject to large uncertainties, from two sources. First,
the next-to-leading calculation is not complete: the gluonic
contributions beyond two-loop order are unknown.
Second, subdominant contributions, such as those
responsible for momentum conservation beyond two-loop
order, are still important at HERA energies.

\vskip 0.7 true cm

\noindent {\bf Acknowledgments.  } We are most grateful to Stefano
Catani for a number of valuable discussions on topics related to
this work. We would like to thank Gavin Salam for help in preparing
Fig.~1. RKE would like to thank the Cavendish and Rutherford-Appleton
Laboratories for their hospitality, and the UK Particle Physics and
Astronomy Research Council for the award of a Senior Fellowship.
FH acknowledges the hospitality of the Theory Group at Fermilab.

\vskip 0.7 true cm

\newpage
{\large \bf References}
\begin{enumerate}

\item \label{Stirl}
      See, for instance, W.J.\ Stirling, in Proceedings of the Aachen
      Conference  ``QCD -- 20 years later", eds. P.M.\ Zerwas and
      H.A.\ Kastrup (World Scientific, Singapore, 1993), p.~387.
\item \label{felt}
      J.\ Feltesse, Saclay preprint Dapnia/SPP 94-35, rapporteur talk
      at the 27th International Conference on High Energy Physics,
      Glasgow, July 1994.
\item \label{revx}
      For recent reviews see, for instance, J.\ Kwiecinski, Krakow Report
      No. 1681/PH, to appear in Proceedings of the Workshop
      ``QCD94", Montpellier, July 1994; \\
      S.\ Catani, Florence preprint DFF 207/6/94, to appear in
      Proceedings of ``Les Rencontres de Physique de la Vallee d'Aoste",
      La Thuile, March 1994.
\item \label{EKL}
      R.K.\ Ellis, Z.\ Kunszt and E.M.\ Levin, \np{420}{517}{94}.
\item \label{BFKL}
      L.N.\ Lipatov, \sj{23}{338}{76}; \\ E.A.\ Kuraev,
      L.N.\ Lipatov and V.S.\ Fadin, Sov.\ Phys.\ JETP  45 (1977) 199; \\
      Ya.\ Balitskii and L.N.\ Lipatov, \sj{28}{822}{78}.
\item \label{FL}
      V.S.\ Fadin and L.N.\ Lipatov, \np{406}{259}{93}.
\item \label{white}
       A.R.\ White, \pl{334}{87}{94}.
\item \label{oldkwie}
      J.\ Kwiecinski, \zp{29}{561}{85};
\item \label{levrysk}
      L.V.\ Gribov, E.M.\ Levin and M.G.\ Ryskin, \prep{100}{1}{83};\\
      E.M.\ Levin and M.G.\ Ryskin, \prep{189}{268}{90}.
\item \label{Dur}
       A.J.\ Askew, J.\ Kwiecinski, A.D.\ Martin and P.J.\ Sutton,
       \pr{47}{3775}{93},  \pr{49}{4402}{94}.
\item \label{HEF}
      S.\ Catani, M.\ Ciafaloni and F.\ Hautmann,
      \pl{242}{97}{90},  \np{366}{135}{91},
      \pl{307}{147}{93}.
\item \label{CHlett}
      S.\ Catani and F.\ Hautmann, \pl{315}{157}{93}.
\item \label{CH}
      S.\ Catani and F.\ Hautmann, \np{427}{475}{94}.
\item \label{uni}
      J.\ Bartels,
      \pl{298}{204}{93}, \zp{60}{471}{93}; \\
      A.H. Mueller, \np{335}{115}{90}, preprint CU-TP-640.
\item \label{GLAP}
      V.N.\ Gribov and L.N.\ Lipatov, \sj{15}{438}{72}, 675; \\
      G.\ Altarelli and G.\ Parisi, \np{126}{298}{77}; \\
      Yu.L.\ Dokshitzer, Sov.\ Phys.\ JETP  46 (1977) 641.
\item \label{2loop}
      E.G.\ Floratos, D.A.\ Ross and C.T.\ Sachrajda,
        \np{152}{493}{79}; \\
      G.\ Curci, W.\ Furmanski and R.\ Petronzio,  \np{175}{27}{80};\\
      W.\ Furmanski and R.\ Petronzio, \pl{97}{437}{80};\\
      E.G.\ Floratos, C.\ Kounnas and R.\ Lacaze,
        \np{192}{417}{81}.
\item \label{EHWprep}
      R.K.\ Ellis, F.\ Hautmann and B.R.\ Webber, Cambridge
      preprint Cavendish-HEP-94/20, in preparation.
\item \label{AEM}
      G.\ Altarelli, R.K.\  Ellis and G.\ Martinelli,
      \np{157}{461}{79}.
\item \label{msbar}
      W.A.\ Bardeen, A.J.\ Buras, D.W.\ Duke and T.\ Muta,
      \pr{18}{3998}{78}.
\item \label{MRS}
      A.D.\ Martin, R.G.\ Roberts and W.J.\ Stirling,
      \pl{306}{145}{93}.
\item \label{E665}
      A.V.\ Kotval, preprint Fermilab-Conf-94/345-E,
      to be published in Proc.\ VIth Rencontres de Blois,
      Blois, June 1994.
\item \label{DATA}
      ZEUS Collaboration, M.\ Derrick et al., \pl{316}{412}{93}; \\
      H1 Collaboration, I.\ Abt et al., \np{407}{515}{93},
      \pl{321}{161}{94}.
\item \label{bryanpino}
      E.M.\ Levin, G.\ Marchesini, M.G.\ Ryskin and B.R.\ Webber,
      \np{357}{167}{91}.
\item \label{zeus93}
      ZEUS Collaboration,  preprint DESY-94-143.
\end{enumerate}


\begin{figcap}

\item \label{1} The BFKL anomalous dimension $\ga_L$ as a function of
the complex moment variable $\w/\abar$:
 a) Real part, b) Imaginary part.

\item \label{2}
a) Gluon and b) quark singlet density in the DIS-scheme
at different values of $Q^2$
($Q^2 = 10, \; 100 \;$ GeV$^2$)
as obtained from
two-loop (dashed), two-loop + L($x$) (dotdashed),
two-loop + NLQ($x$) (solid) evolution
of  flat input distributions (MRSD0$^{\prime}$) at $Q^2_0 = 4$ GeV$^2$.
The dotted curves correspond to NLQ($x$) resummation with the
harder model for momentum conservation described in the text.

\item \label{3} Resummed predictions for the structure function $F_2$.
The short-dashed
curve is the one-loop fixed-order  prediction.  The other curves are
as in Fig.~2.
\end{figcap}
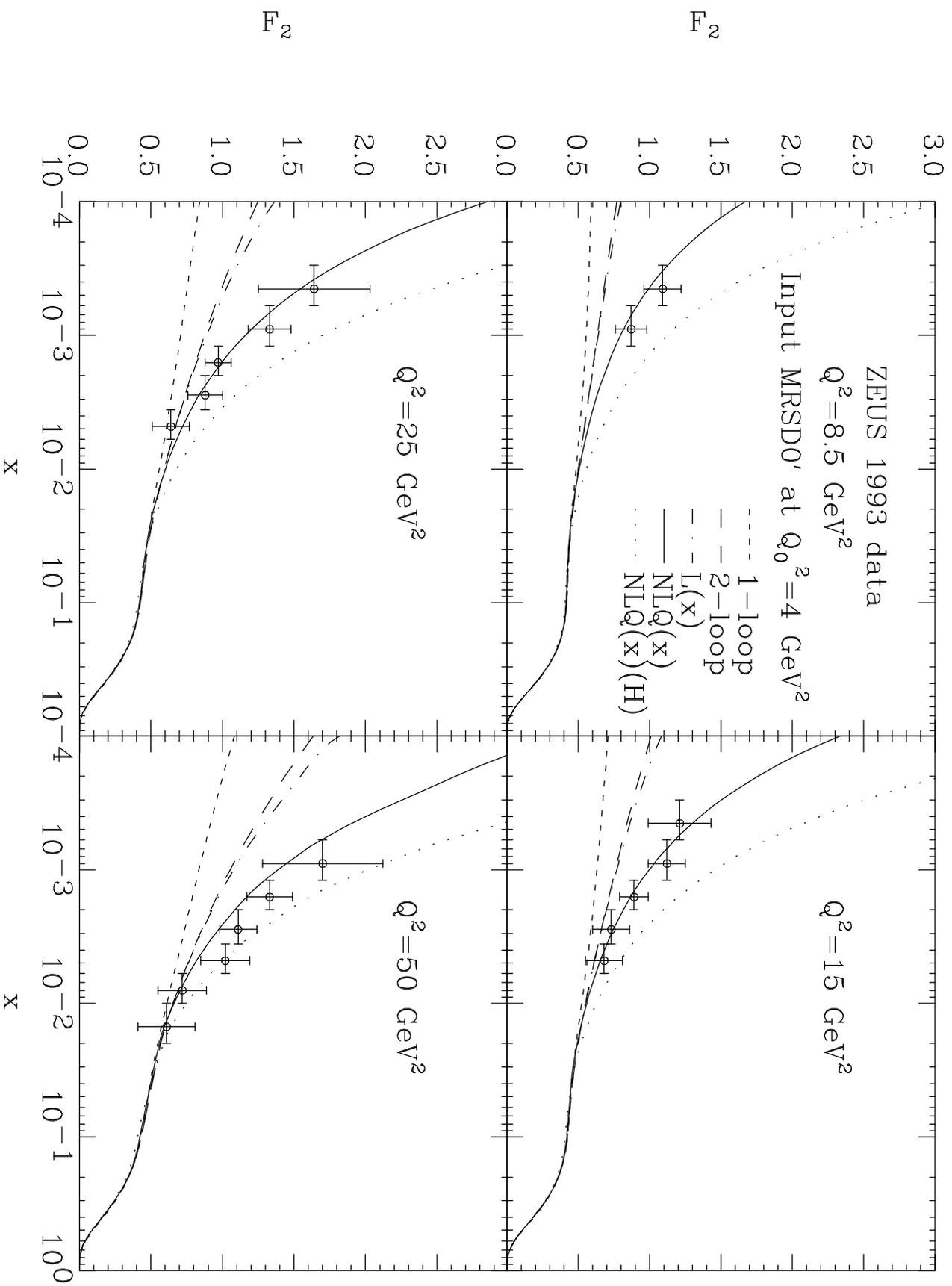
\end{document}